\documentclass{article}
\usepackage[T1]{fontenc} 
\usepackage[utf8]{inputenc} 
\usepackage{ismir,amsmath,cite,url}
\usepackage{graphicx}
\usepackage{tabularx,booktabs}
\usepackage{color}
\usepackage{kotex}
\usepackage{array}
\usepackage{amsfonts}

\newcommand{\norm}[1]{\left\lVert#1\right\rVert}
\newcolumntype{P}[1]{>{\centering\arraybackslash}p{#1}}
\newcommand\blfootnote[1]{%
  \begingroup
  \renewcommand\thefootnote{}\footnote{#1}%
  \addtocounter{footnote}{-1}%
  \endgroup
}

\title{Audio query-based music source separation}

\multauthor
{\textsuperscript{*}Jie Hwan Lee$^{1,2}$ \hspace{1cm} \textsuperscript{*}Hyeong-Seok Choi$^{1,2}$ \hspace{1cm} Kyogu Lee$^{1,2}$} {\\
  $^1$Music \& Audio Research Group, Seoul National University, Korea\\
  $^2$Center for Super Intelligence, Seoul National University, Korea\\
{\tt \{wiswisbus, kekepa15, kglee\}@snu.ac.kr}
}
\def\authorname{Jie Hwan Lee, Hyeong-Seok Choi, Kyogu Lee}

\begin{document}
\maketitle
\begin{abstract}
In recent years, music source separation has been one of the most intensively studied research areas in music information retrieval. Improvements in deep learning lead to a big progress in music source separation performance. However, most of the previous studies are restricted to separating a few limited number of sources, such as vocals, drums, bass, and other. In this study, we propose a network for audio query-based music source separation that can explicitly encode the source information from a query signal regardless of the number and/or kind of target signals. The proposed method consists of a Query-net and a Separator: given a query and a mixture, the Query-net encodes the query into the latent space, and the Separator estimates masks conditioned by the latent vector, which is then applied to the mixture for separation. The Separator can also generate masks using the latent vector from the training samples, allowing separation in the absence of a query. We evaluate our method on the MUSDB18 dataset, and experimental results show that the proposed method can separate multiple sources with a single network. In addition, through further investigation of the latent space we demonstrate that our method can generate continuous outputs via latent vector interpolation.

\end{abstract}
\section{Introduction}\label{sec:introduction}
\blfootnote{*these authors contributed equally}
Music source separation, isolating the signals of certain instruments from a mixture, has been intensively studied in recent years. Due to the improvements in deep learning techniques, various approaches using deep learning for music source separation have been introduced.
However, most of the previous studies are mainly focused on improving music source separation performances, not the range of separable sources.
To tackle this problem, a few studies have tried to separate the fixed number of sources of interest by conditioning one-hot label in the deep learning network \cite{seetharaman2018class,slizovskaia2018end}.

While being the most straight-forward approach, we argue that such an approach is not a proper way to deal with the \textit{outliers}
when the generic and broadly defined class labels are the only available data at hand \cite{SiSEC16, musdb18}.
To understand this situation more concretely, let us consider the  mismatched situations where the target source is classified into a certain generic class but still somewhat far from the general characteristics of that broadly defined generic class.
For example, consider the situation where we desire to separate `distorted singing voice' or `acoustic guitar' sources. 
In these cases, we can imagine that the performance can be boosted if we were to have more fine-grained labels such as `distorted singing voice' or `acoustic guitar' rather than generic classes such as `vocals' or `guitar'.
One of the simplest ad-hoc solutions, therefore, can be manually annotating such \textit{outliers} based on the music instrument ontology and conditioning those new classes into the deep learning network.
Unfortunately, manually annotating an audio signal has limitation in many aspects. 
First, labeling an audio itself is costly.
Second, given the same audio samples, the number of samples per class is reduced, hence it is likely that the separation performance degrades.
Third, such a method is not scalable to new outlier samples, and is thus limited.

\begin{figure}[h]
\centering
{\includegraphics[scale=.4]{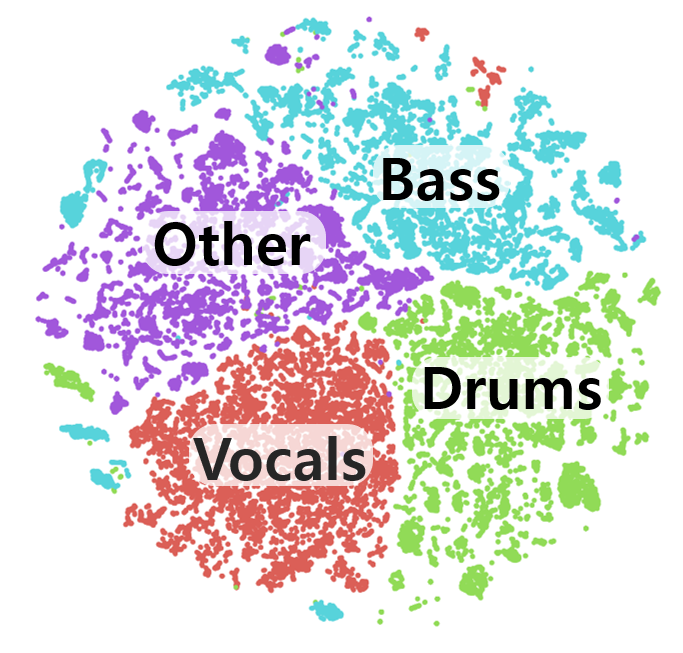}} \\
\caption{t-SNE visualization \cite{maaten2008visualizing} of encoded latent vectors of the test dataset in MUSDB18.
Without any classification loss, the Query-net is trained to output latent vectors that provide useful information about various instruments.
It is observable that the latent vectors from the same class are clustered in the latent space while not being identical.
}
\label{fig:tsne}
\end{figure}

To deal with these problems, in this paper, a novel audio query-based music source separation framework is proposed.
The main idea is to directly compress the diverse audio samples into latent vectors -- using the so-called \textit{Query-net} -- so that the audio samples can be mapped into non-identical points even when the samples are from the same class as illustrated in Fig. \ref{fig:tsne}.
The encoded latent vector is then fed into a separation network to output a source whose characteristics is similar to the audio sample taken into the Query-net.
The proposed framework is scalable as the Query-net is able to encode an unseen singing voice or instrument sound into the continuous latent space.
This property allows many useful utilities as follows.
First, it is capable of separating various number of sources with a single network.
Second, we can expect an increase in separation performance especially when the characteristics of the target source in the mixture is considered far from the given generic class since the user can manually select and encode the held-out sound sample that is deemed similar to the target signal.
Third, it allows the natural control of the output of the separation network by interpolating the latent vectors in the continuous latent space.

To demonstrate the usefulness of the proposed method, we show various experiments using the MUSDB18 dataset \cite{musdb18}.
The experiments show that the output of the separation network is highly dependent on the latent vector which allows smooth transition in signal level by controlling and interpolating the latent vectors.
Also, we show that the proposed method becomes especially useful when the target source of interest is far from the general characteristic of coarsely defined sound class.
Finally, we show that the proposed method can be even automated by iteratively encoding the separation output.

\section{Related work}
In this section, we first introduce previous music source separation studies that tried to separate mixture into multiple sound classes.
One of the most basic ways is to estimate several separation masks with a single model. In \cite{park2018music}, they tried to separate four sources with one stacked hourglass model \cite{newell2016stacked}. While they showed a competitive results the method is not flexible as the model requires a fixed number of output.
Next, \cite{slizovskaia2018end} introduced a one-hot label conditioning approach and showed that their proposed method is capable of separating multiple sources. This method is more flexible than the aforementioned model but the model does not assume latent space, therefore, is not capable of manipulating output other than conditioning the one-hot label.
Finally, \cite{seetharaman2018class} showed that they can embed each time-frequency bin of the mixture into a high-dimensional space using deep clustering \cite{hershey2016deep} approach.
However, this approach still has a limitation in that the model is not capable of encoding the audio signal directly into the latent space.
Apart from the music source separation studies, \cite{WangCSCYQY18} suggested a speaker-dependent speech separation method by incorporating a lstm-based anchor vector encoder which enables direct encoding of audio signal into a latent space. Using this technique, they showed that the proposed method can cluster the time-frequency bin embeddings that are close to the anchor vector in the latent space.

\begin{figure}[t]
\centering
{\includegraphics[scale=.35]{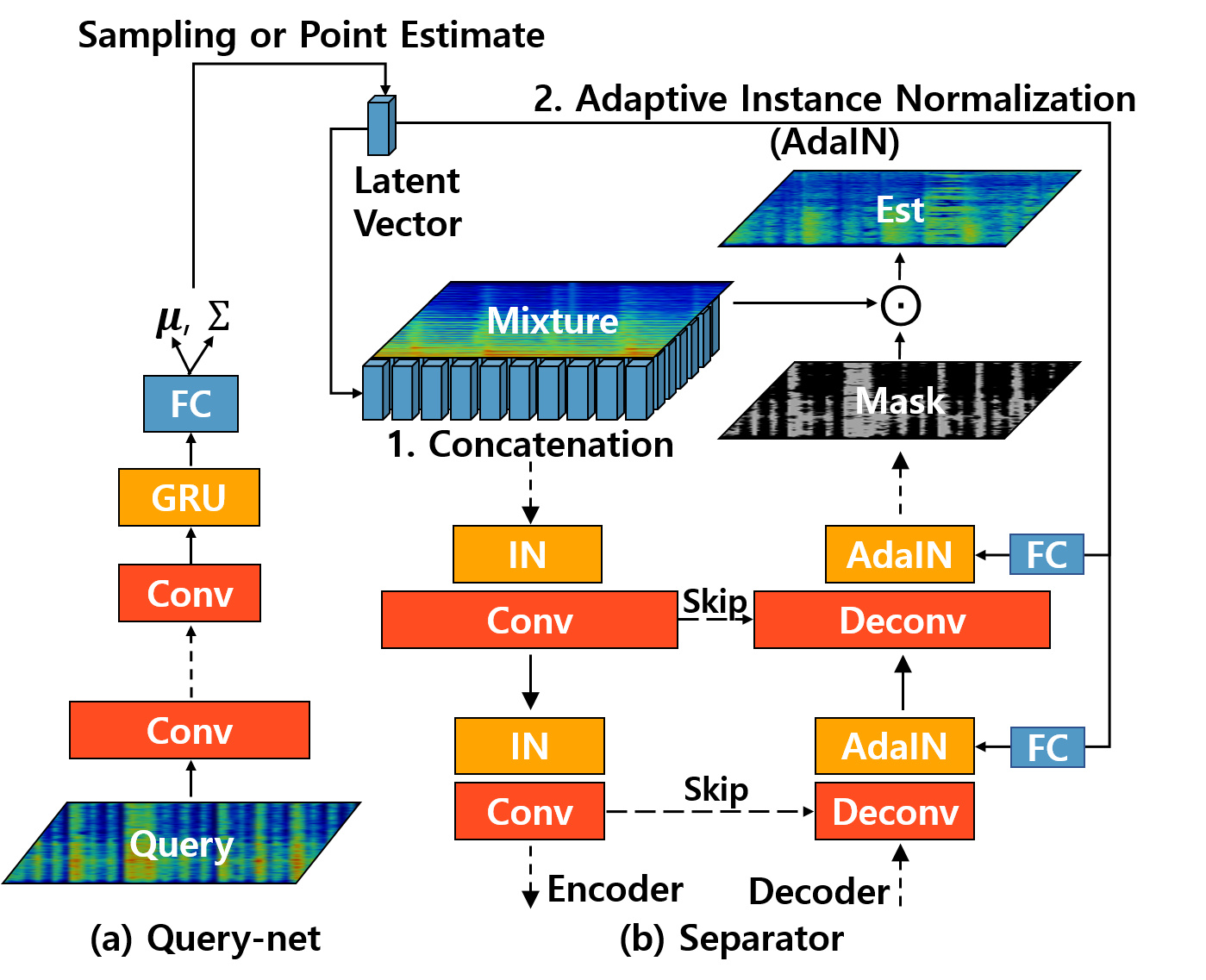}} \\
\caption{Illustration of the (a) Query-net and (b) Separator. The Query-net encodes the query into the latent vector and it is passed into the Separator by two methods. 1. Concatenation: The latent vector is concatenated with mixture spectrogram by tiling the latent vector along the spatial dimension. 2. AdaIN: Adaptive instance normalization is used in every layer of decoder part. 
}
\label{fig:separator}
\end{figure}

\section{Proposed Method}\label{sec:proposed_method}

\subsection{Query-based Source Separation}\label{subsec:architecture}
The proposed framework is composed of two deep learning networks, Query-net $\mathbb{Q}(\cdot)$ and Separator $\mathbb{S}(\cdot)$.
While most of the previous studies typically use $\mathbb{S}$ to extract a single class source from a mixture, we aim to separate the mixture by manipulating the additional input signal, a query.
By doing so, we can expect to have a control over the mixture just by choosing a different query input which can be done either manually by the user or automatically by the system.
Hence, the query signal is expected to be sampled from a similar sound class to the target signal within the mixture, but does not have to be identical.
To achieve this, $\mathbb{Q}$ directly encodes the query audio signal into a latent vector so that we can control the output of $\mathbb{S}$ by manipulating the latent space.

$\mathbb{Q}$ is composed of 6 strided-convolutional layers followed by gated recurrent unit (GRU) layer.
The stack of strided-convolutional layers are used to extract local features from the given query signal. Then, the extracted features are reshaped by stacking each feature map along the frequency axis. Finally, the reshaped tensor is passed into GRU and the last state of the GRU is used as a summary of the query signal.
As we would like the encoded latent vector to have a meaningful high-level information, 
we designed $\mathbb{Q}$ to map the query into a small enough dimension compared to the dimension of the query signal.
After the audio query has been encoded, the summarized information is passed into $\mathbb{S}$. 

$\mathbb{S}$ is a U-Net \cite{ronneberger2015u} based network which has proven its effectiveness in many source separation studies \cite{jansson2017singing, takahashi2017multi, takahashi2018mmdenselstm, StollerED18, park2018music}.
It is a convolutional encoder and decoder with skip-connections between the layers.
$\mathbb{S}$ takes the mixture signal and estimate a sigmoid mask to separate the mixture into a source given the summarized information of query from $\mathbb{Q}$.
To effectively pass the summary of the query signal to $\mathbb{S}$, we applied two methods.
First, we simply concatenated the latent vector along the channel dimension of the input mixture spectrogram expecting the summarized information to be delivered from the start.
Second, we used the adaptive instance normalization (AdaIN) technique in the decoding stage of $\mathbb{S}$, which is proven to be effective in many studies for conditioning latent vectors \cite{karras2018style, huang2017arbitrary}.
AdaIN is simply done by applying two steps on each output $\mathbf{x}$ of the convolutional layer (before activation) of the decoder part of $\mathbb{S}$.
First, each $i$-th feature map $\mathbf{x}_{i}$ is normalized using instance normalization technique \cite{huang2017arbitrary}.
Second, affine transformation is applied to the normalized feature map using learned scale and bias parameters which transforms encoded query vector $\mathbf{z}$ into $\mathbf{y}_s$ and $\mathbf{y}_i$ respectively as follows, $\mathbf{y}_{s}=W_{s}^T \mathbf{z}$,  $\mathbf{y}_{b}=W_{b}^T \mathbf{z}$, where $W_{s}$ and $W_{b}$ denote the trainable parameters,
\begin{equation}
    \text{AdaIN}(\mathbf{x}_{i}, \mathbf{y})=\mathbf{y}_{s,i}\cdot(\frac{\mathbf{x}_{i}-\mu(\mathbf{x}_{i})}{\sigma(\mathbf{x}_i)})+\mathbf{y}_{b,i}.
\end{equation}
The overall framework of the proposed method is illustrated in Fig. \ref{fig:separator}.

\subsection{Training}\label{subsec:training}
\subsubsection{Data Sampling}\label{subsubsec:data_sampling}
We first describe how the mixture and target source are selected throughout the training phase.

Let, $v_{i}$ be the single source sampled from $i$-th source class, where $i\in \{1,2,3...,K\}$ and $K$ denote the total number of source classes. 
We split the classes into two groups by randomly assigning each source class into group $T$ (Target) and $R$ (Rest) without replacement until every class is assigned to one of the two groups.
Next, we multiply binary value $\alpha_i$ to the $v_i$, where $\alpha_i$ being sampled from the Bernoulli distribution, $\alpha_i \sim Bernoulli(0.5)$.
This was done to make sure that there are not too many sources included in the mixture.
After then, as a data augmentation strategy \cite{uhlich2017improving}, we scale each source by multiplying a value $\beta_i$ to source $v_i$, where $\beta_i$ is sampled from the Uniform distribution, $\beta_i \sim \mathcal{U}$[0.25, 1.25].
Finally, the sources in each group is added to form two waveforms $s_T$ and $s_R$ and the mixture $m$ is constructed as the linear sum of $s_T$ and $s_R$ as follows,
\begin{equation}
\label{eq:data}
\begin{split}
    m = s_T + s_R = \sum_{i \in T} (\beta_i \cdot \alpha_i \cdot v_i) + \sum_{j \in R} (\beta_j \cdot \alpha_j \cdot v_j).
\end{split}
\end{equation}
As we used magnitude spectrogram as input of the modules, $m$, $s_T$, and $s_R$ are transformed into short-time-Fourier-transform (STFT) domain, which we denote in capital letter $M$, $S_T$ and $S_{R}$, respectively.
Note that, we do not assume any musicality of mixture signal, hence each class is sampled from arbitrary mixture tracks.

\subsubsection{cVAE with Latent Regressor}\label{subsubsec:latent_regressor}
To design the proposed framework, we borrow the formulation of conditional variational autoencoder (cVAE).
While the latent vector $\mathbf{z}$ can be deterministically encoded into the latent space, in cVAE framework, $\mathbf{z}$ is instead sampled from the Gaussian distribution, where the parameters of the distribution (mean and variance) are estimated from $\mathbb{Q}$.
Then, $\mathbb{S}$ is used to reconstruct $S_T$ given $M$ and $\mathbf{z} \sim \mathbb{Q}(S_T)$.
This is ensured by one of the two objectives of cVAE, namely, reconstruction loss $\mathcal{L}_{R}$.
The purpose of $\mathcal{L}_{R}$ is to guarantee that the output of $\mathbb{S}$ is dependent on the encoded latent vector as follows,
\begin{equation}
\label{eq:recon}
\begin{split}
    \mathcal{L}_{R} = \mathbb{E}_{S_T \sim p(S_T), \, M \sim p(M),  \, \mathbf{z} \sim \mathbb{Q}(S_T)} [\norm{ S_T -  \mathbb{S}(M, \mathbf{z}) }]_1.
\end{split}
\end{equation}
Note that, in training phase, the latent vector $\mathbf{z}$ is sampled using re-parameterization trick to allow backpropagation in training phase \cite{KingmaW13}.

Next, KL-divergence loss is used to make the distribution of $\mathbf{z}$ be close to the Gaussian distribution $\mathcal{N}(\mathbf{0},I)$ to guarantee a sampling at test time.
\begin{equation}
\label{eq:kl}
\begin{split}
    \mathcal{L}_{\mathrm{KL}}=\mathbb{E}_{S_T \sim p(S_T)}\left[\mathcal{D}_{\mathbf{K L}}(\mathbb{Q}(S_T) \| \mathcal{N}(\mathbf{0}, I))\right]
\end{split}
\end{equation}

Apart from cVAE framework, we also adopted latent regressor used in \cite{zhu2017toward} to enforce the output of $\mathbb{S}$ to be more dependent on the latent vector.
First, a random vector $z$ is drawn from the prior Gaussian distribution $\mathcal{N}(\mathbf{0},I)$ and passed to $\mathbb{S}$. 
Then, $\mathbb{S}$ produces a reasonable output reflecting the information in the random vector.
Finally, $\mathbb{Q}$ is reused to restore the random vector from the output from $\mathbb{S}$.
Note that, unlike Eq. \ref{eq:recon} and \ref{eq:kl}, only the mean values ($\mathbf{\mu}$) are taken from $\mathbb{Q}$ as a point estimate of $\mathbf{z}$.
\begin{equation}
\label{eq:regressor}
    \mathcal{L}_{latent}=\mathbb{E}_{M \sim p(M), \, \mathbf{z} \sim p(\mathbf{z})}\|\mathbf{z}-\mathbb{Q}(\mathbb{S}(M, \mathbf{z}))\|_{1}
\end{equation}

Finally, the total loss can be written as follows, 
\begin{equation}
\label{eq:total}
    \mathcal{L}_{Total}=\lambda_{R}\mathcal{L}_{R} + \lambda_{\mathrm{KL}}\mathcal{L}_{\mathrm{KL}} + \lambda_{latent}\mathcal{L}_{latent}.
\end{equation}

\subsection{Test} \label{subsec:test}
During the training phase, $\mathbb{S}$ was trained to separate the target source by using the target source as a query as in Eq. \ref{eq:recon}.
In the test phase, however, the target source to be separated from the mixture is unknown.
Hence, the target source and query can no longer be the same.
Nevertheless, since we designed the output dimension of $\mathbb{Q}$ to be small enough, the latent vector $\mathbf{z}$ is trained to have a high-level information such as instrument class.
In the test phase, therefore, we can utilize this property in many ways.
For example, when the user wants to separate a specific source in the mixture, it is possible to collect a small amount of audio samples that have similar characteristics but not exactly the same to the source of interest.
Then, the user can extract that specific source by feeding the collected audio samples into the Query-net and passing the summarized information to the Separator.

Apart from the query dependent approach, we can also take the average of latent vectors of each source class in the training set and use it as a representative latent vector that reflects the general characteristics of a single class. 

\section{Experiment}\label{sec:experiment}

\subsection{Dataset}\label{subsec:body}

We trained our network with the MUSDB18 dataset. 
The dataset consists of 100 tracks for training set and 50 tracks for test set and each track is recorded in 44.1kHz, stereo format. 
The dataset provides the mixture and coarsely defined labels for sources, namely, `vocals', `drums', `bass' and `other'.
The class `other' includes every instrument other than `vocals', `drums' and `bass', providing the most coarsely defined class.
We resampled the audio to 22050Hz and divided each track into 3-second segments. Magnitude spectrogram was obtained by applying STFT with a window size 1024 and 75\% overlap. To restore the audio from the output, Inverse STFT is applied using the phase of the mixture. 
We evaluated our method on the test set of MUSDB18 using the official museval package\footnote{https://sigsep.github.io/sigsep-mus-eval} which computes signal-to-distortion ratio (SDR) as a quantitative measurement.

\subsection{Experiment Details}

The followings are the experimental details of our method. 
$\mathbb{Q}$ consists of 6 strided-convolutional layers with 4 $\times$ 4 filter size and the number of output channels for each layer is 32, 32, 64, 64, 128 and 128, respectively. Every strided-convolutional layer has the stride size of 2 along the frequency axis and only second, fourth and sixth layers have a stride size of 2 along the time axis. After every convolutional operation, we used instance normalization and relu. We used GRU with 128 units. 
The length of the query segment was fixed to 3-second in every experiment.
For $\mathbb{S}$, the encoder part consists of 9 strided-convolutional layers and the decoder part consists of the same number of strided-deconvolutional layers, with a filter size of 4 $\times$ 4. 
The number of output channels for first, second, and third layer is 64, 128, 256, respectively, and 512 for the rest of the layers. Every layer has stride size of 2 along the frequency axis. And stride size along the time axis is set to 2 for every layer except the first layer of the encoder and the last layer of the decoder.

The dimension of the latent vector was set to 32 and the batch size was set to 5. The coefficients in Eq.\ref{eq:total} were set to $\lambda_{R} = 10$, $\lambda_{\mathrm{KL}} = 0.01$, $\lambda_{latent} = 0.5$.  The initial learning rate was set to 0.0002 and after 200000 iterations the rate was decreased to $5\times10^{-6}$ for every 10000-iteration. We used Adam optimizer with $\beta_1 = 0.5$,  $\beta_2 = 0.999$.

\begin{figure}[t]
 \centerline{
 \includegraphics[width=\columnwidth]{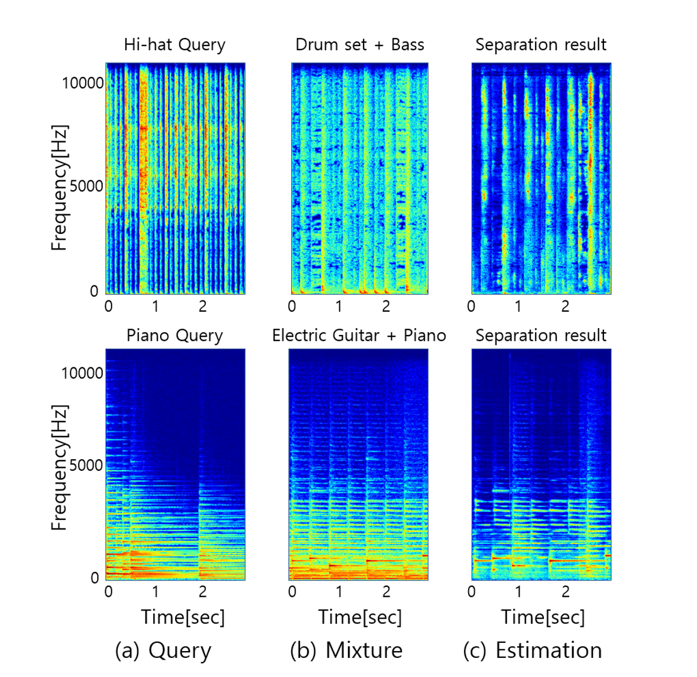}}
 \caption{Results of manually targeting specific sound sources. The first row show the separation results of hi-hat from the mixture of hi-hat, kick drum and bass.
 The second row shows the separation results of piano from the mixture of electric guitar and piano. It is worth noting that the network was never trained to only separate a hi-hat component from `drum' class nor piano from `other' class.
}
 \label{fig:specific_instrument} 
\end{figure}

\begin{figure}[t]
 \centerline{
 \includegraphics[width=\columnwidth]{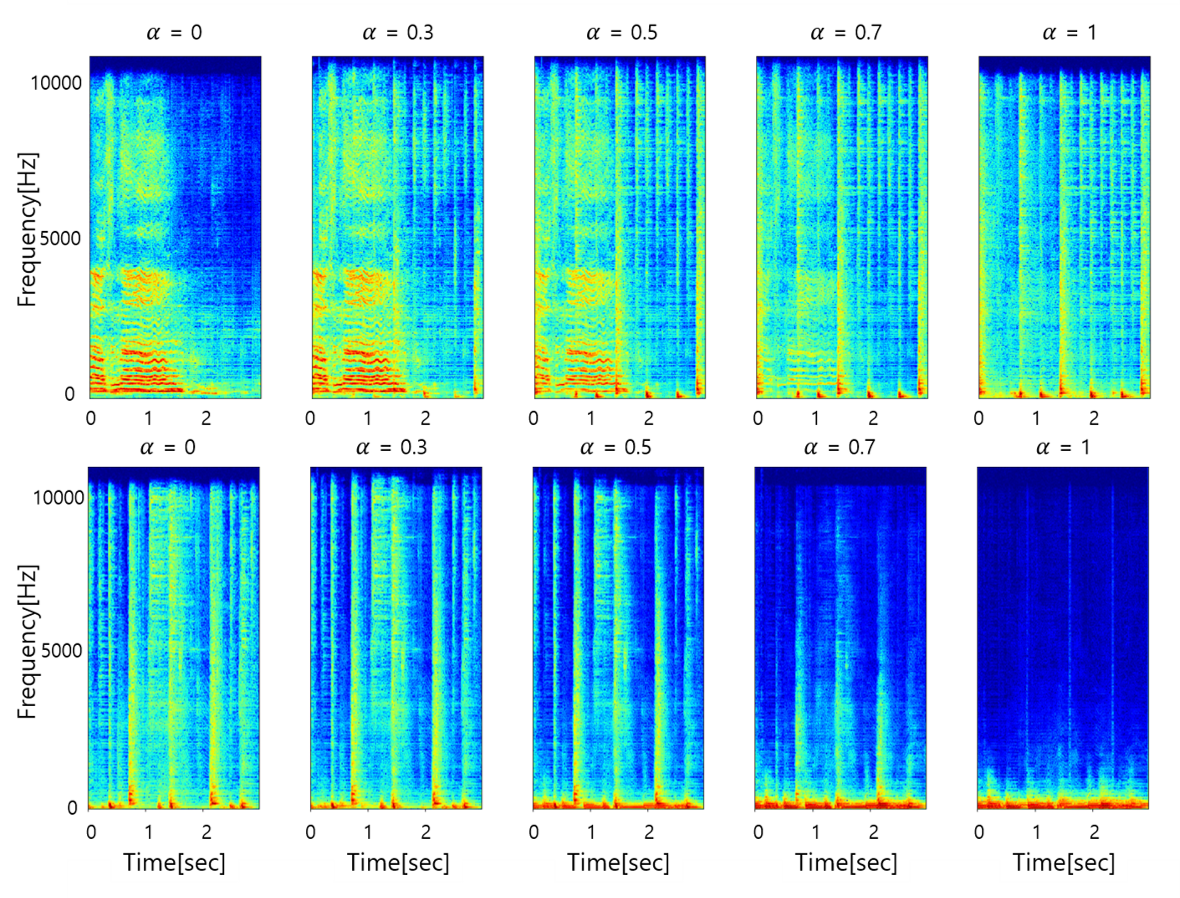}}
 \caption{Results of the mean vector interpolation.
 The first row shows the interpolation results between vocals and drums. 
 The second row shows the interpolation results between drums and bass.
}
 \label{fig:Latent interpolation} 
\end{figure}

\subsection{Manually Targeting a Specific Sound Source}

To validate that our method captures the characteristics of the audio given in the query and separates them accordingly, we conducted an experiment of separating specific instruments. As shown in Fig. \ref{fig:specific_instrument}, an audio query of hi-hat and piano were given to the mixtures of (hi-hat $+$ kick drum $+$ bass) and (piano $+$ electric guitar). Queries and mixtures were not from the train set, and both queries were not sampled from the mixture. We can observe in the hi-hat separation result that the kick drums and the bass which lie in the low-frequency band were mostly removed while broadband components of hi-hat remained. The result of piano separation is not as clear as in the case of hi-hat, but we can see the guitar was removed considerably. 

The noticeable fact is that we trained our method only with the MUSDB18 dataset, which has no hierarchical class label information besides the coarsely defined labels of sources such as `vocals', `drums', `bass' and `other'. 
Under the definition of class in the dataset, hi-hat and kick drum are grouped into `drums', and piano and electric guitar into `other'. 
Although our method was never trained to separate the subclass from the mixture, it was able to separate hi-hat and piano from the mixture, which can be referred to as a zero-shot separation.
These results indicate the proposed method can be well applied for audio query-based separation.

\subsection{Latent Interpolation}

Furthermore, we conducted a latent interpolation experiment using the mean vector of each source.
The mean vector of each source was computed by averaging the latent vectors of each source in the training set, $\mathbf{z}_c = \frac{1}{N_c} \sum_{i}{\mathbb{Q}(S_{c,i})}$, where $S_{c,i}$ denotes $i$-th 3-second magnitude spectrogram in the sound class $c$ and $N_c$ denotes the number of segments in class $c$.

For the interpolation method, we used the spherical linear interpolation (${Slerp}$) introduced in \cite{white2016sampling},
\begin{equation}
\mathit{Slerp}({\mathbf{z}}_{1},{\mathbf{z}}_{2};\alpha) = \frac{\sin(1 - \alpha)\theta   }{\sin\theta}{\mathbf{z}}_{1} + \frac{\sin\alpha\theta}{\sin\theta}{\mathbf{z}}_{2},
\end{equation}
where $\alpha$ denotes the weight of interpolation and $\theta$ denotes the angle between ${\mathbf{z}}_{1}$ and ${\mathbf{z}}_{2}$.
As shown in Fig. \ref{fig:Latent interpolation}, we interpolated between the mean vector of sound sources, drums ($\mathbf{z}_{drums}$) $\rightarrow$ bass ($\mathbf{z}_{bass}$) and vocals ($\mathbf{z}_{vocals}$) $\rightarrow$ drums ($\mathbf{z}_{drums}$). We can see the ratio of separated instruments changes as the weight $\alpha$ changes. These experimental results show that our method can generate continuous outputs just by manipulating a latent space.

\subsection{Effects of Latent Vector on Performance}

\begin{figure}[h]
\centering
{\includegraphics[scale=.4]{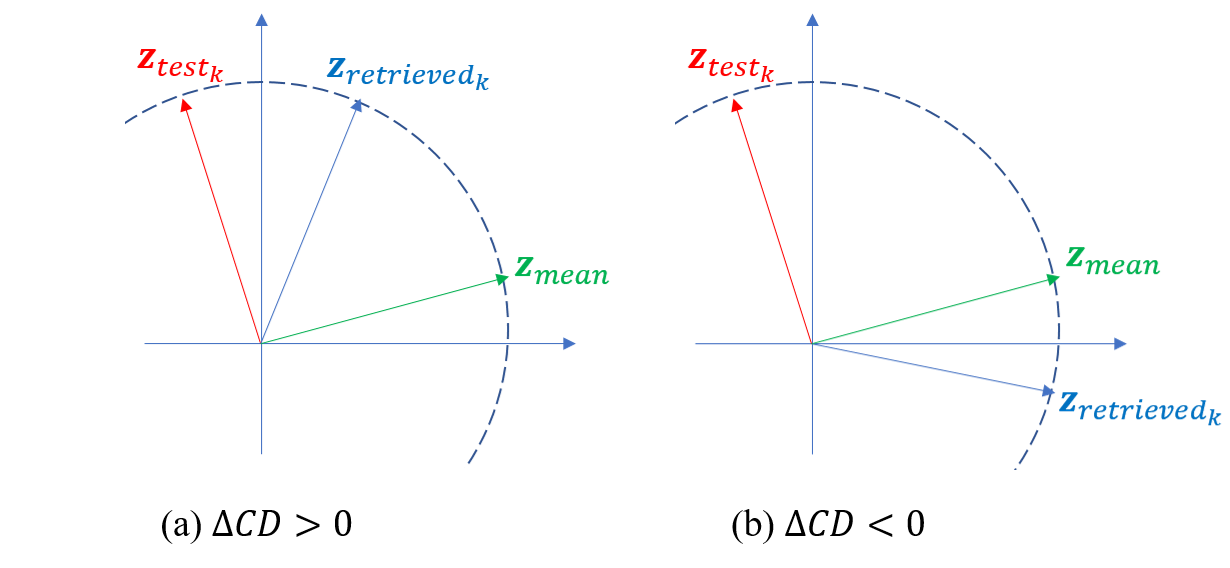}} \\
\caption{Illustration of two $\Delta CD$ cases. 
(a) shows the positive $\Delta CD$ case where we assume that the performance should be improved. (b) shows the negative $\Delta CD$ case where we assume that the performance should be worsened.
}
\label{fig:CD}
\end{figure}

\begin{figure}
 \centerline{
 \includegraphics[width=\columnwidth]{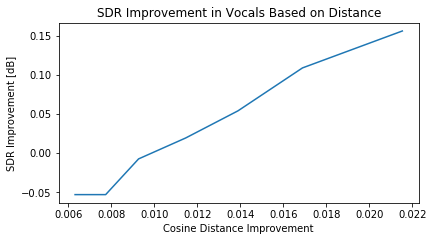}}
 \caption{The relationship between SDR improvement ($\Delta$SDR) and cosine distance difference ($\Delta CD$) in vocal tracks.
}
 \label{fig:SDRimprovement} 
\end{figure}
This subsection investigates the performance improvement varying the latent vector and see in which situation we can achieve a performance improvement.
For the experiment, we first obtained the mean vector of each vocal track from the entire dataset as follows, $\mathbf{z}_i = \frac{1}{N_i}\sum_{j} \mathbb{Q}(S_{i, j})$, where $i$ denotes a $i$-th vocal track, $j$ denotes a $j$-th segment in $i$-th vocal track, and $N_i$ denotes the number of segments in $i$-th vocal track.
Then, we obtained the mean vector, $\mathbf{z}_{mean} = \frac{1}{100} \sum_{i \in training} \mathbf{z}_i$, of vocal tracks from training set. 
Finally, we retrieved the latent vector of certain vocal track $\mathbf{z}_{ret_k}$ from the training set which has the closest cosine distance ($CD$) from $k$-th test vocal track $\mathbf{z}_{test_k}=\mathbf{z}_{k}, \, k \in test$ as follows, 
\begin{equation}
\tilde{k}=\underset{i\in training}{\arg\min} \, CD(\mathbf{z}_i, \mathbf{z}_{test_k}),\,\,\, \mathbf{z}_{ret_k}=\, \mathbf{z}_{\tilde{k}}.
\end{equation}

We compare the performance of two cases where the goal is to separate a $k$-th vocal track from test set.
The first case is to use $\mathbf{z}_{mean}$ to separate a target source, $\hat{S}_{mean} = \mathbb{S}(M, \mathbf{z}_{mean})$.
The second case is to use $\mathbf{z}_{ret_k}$ to separate a target source, $\hat{S}_{ret_k} = \mathbb{S}(M, \mathbf{z}_{ret_k})$.
We defined performance improvement in terms of SDR as follows,
\begin{equation}
\label{eq:delta_sdr}
\Delta\text{SDR}=\text{SDR}(S_{GT_k},\hat{S}_{ret_k})-\text{SDR}(S_{GT_k},\hat{S}_{mean}),
\end{equation}
 where $S_{GT_k}$ denotes $k$-th ground truth vocal track from test set.
To measure the distance between latent vectors we used cosine distance ($CD(\mathbf{z}_1, \mathbf{z}_2) = 1- (\mathbf{z}_1 / \norm{\mathbf{z}_1}_2) \cdot (\mathbf{z}_2 / \norm{\mathbf{z}_2}_2)$) and defined cosine distance difference between ($\mathbf{z}_{test_k}, \mathbf{z}_{ret_k}$) and ($\mathbf{z}_{test_k}, \mathbf{z}_{mean}$) as follows,
\begin{equation}
\label{eq:delta_cosinedistance}
\Delta CD =
 CD(\mathbf{z}_{test_k}, \mathbf{z}_{mean}) - CD(\mathbf{z}_{test_k}, \mathbf{z}_{ret_k}).
\end{equation}

Fig. \ref{fig:CD} illustrates two possible cases of using $\mathbf{z}_{ret_k}$.
(a) shows the positive $\Delta CD$ case where we assume to induce positive effect on performance improvement ($\Delta \text{SDR} > 0$).
In this case, we expect the performance to be improved since $\mathbf{z}_{ret_k}$ is expected to contain information close to $\mathbf{z}_{test_k}$ compared to $\mathbf{z}_{mean}$.
(b) shows the negative $\Delta CD$ case where we assume to induce negative effect on performance improvement ($\Delta \text{SDR} < 0$).
In this case, we expect the performance to be worsened as the system could not retrieve a $\mathbf{z}_{ret_k}$ that is close enough to $\mathbf{z}_{test_k}$.
To empirically prove our assumption, we show the relationship between $\Delta$SDR and $\Delta CD$ in Fig. \ref{fig:SDRimprovement}.
We can observe that the closer the vector gets to the targeted ground-truth vector, the larger the performance gain becomes, therefore reinforcing our assumption that better performances can be achieved if we can obtain closer latent vectors to the target latent vector.


\subsection{Iterative Method}

In this subsection, we seek a performance improvement by automating the query-based framework in an iterative way, which we refer to as an iterative method.
The iterative method is done as follows.
First, we separate the target source using the mean vector of certain sound class $\mathbf{z}_{mean}$. Then, we re-encode the separated source into a latent space expecting the re-encoded latent vector to be closer to the target latent vector. Finally, we separate the target source using the re-encoded latent vector.
We verify the effect of the proposed iterative method and show that it can be helpful under the harsh condition where 
the target sources are far from generic class.
The results (Single step $\rightarrow$ Iterative) are as follows, `vocals': $4.84 \rightarrow \mathbf{4.90}$, `drums': $4.31 \rightarrow \mathbf{4.34}$, `bass': $\mathbf{3.11} \rightarrow$ 3.09, and `other': $2.97 \rightarrow \mathbf{3.16}$. We can see the iterative method noticeably improves the performance in `vocals' and `other'. On the other hand, the differences are not significant in drums and bass. 

We looked into the tracks which gained significant improvement in terms of SDR in vocals. ‘Timboz - Pony’ and ‘Hollow Ground - Ill Fate’ gained more than 0.5dB in SDR through the iterative method.
We found the results intuitive as the vocals in the two songs feature a growling technique from heavy metal genres, which can be considered distant from the general characteristics of vocals.


\begin{figure}[h]
\centering
{\includegraphics[scale=.4]{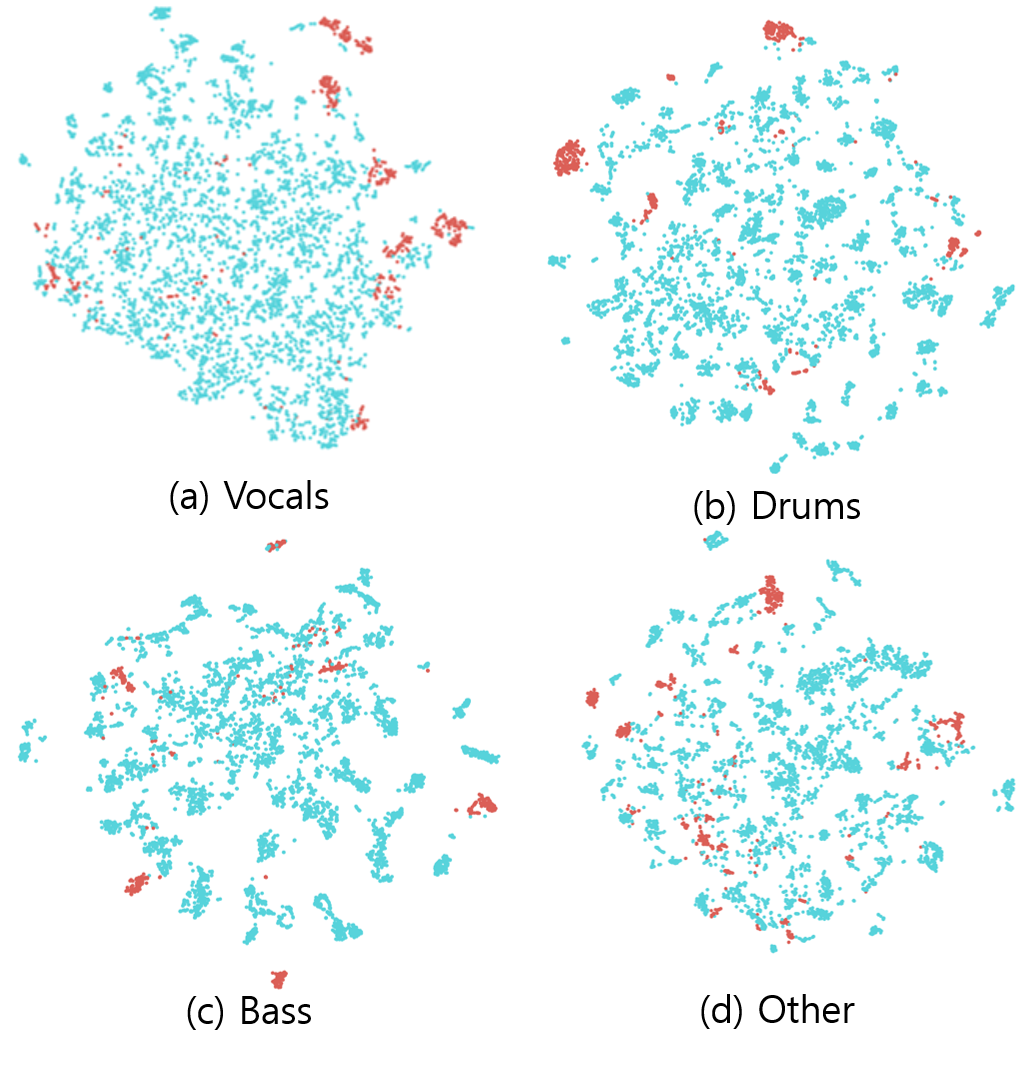}} \\
\caption{
t-SNE visualization of encoded latent vectors from each source in the test dataset. Red points denote the vectors of the tracks which gained more than 0.4dB in terms of SDR by the iterative method.
}
\label{fig:tsne_vocal}
\end{figure}
p
To verify our assumptions, we divided each source of the test set into segments and converted them into latent vectors. 
We divided the encoded vectors into two groups, the ones which gained more than 0.4dB in terms of SDR by the iterative method and the ones did not. 
Then, we visualized the encoded vectors using t-SNE (results shown in Fig. \ref{fig:tsne_vocal}). The red dots in Fig. \ref{fig:tsne_vocal} represent the latent vector from the group that showed significant SDR improvement more than 0.4dB. 
Although some vectors lie around the center, most of them are located far from the center. These vectors can be inferred as outliers and the results show that our iterative method is effective when it comes to separating the sources of distinctive characteristics.

\begin{table}
\begin{center}
\begin{tabular}{|P{2cm}|P{1cm}|P{1cm}|P{1cm}|P{1cm}|}
\hline
     & Vocals        & Drums         & Bass          & Other         \\ \hline
STL2\cite{StollerED18} & 3.25          & 4.22          & 3.21          & 2.25          \\
WK\cite{weninger2014discriminatively}   & 3.76          & 4.00          & 2.94          & 2.43          \\
RGT1\cite{roma2018improving} & 3.85          & 3.44          & 2.70          & 2.63          \\
JY3\cite{liu2018denoising}  & 5.74          & 4.66          & 3.67          & 3.40          \\
UHL2\cite{uhlich2017improving} & 5.93          & 5.92          & 5.03          & \textbf{4.19}          \\
TAK1\cite{takahashi2018mmdenselstm} & \textbf{6.60} & \textbf{6.43} & \textbf{5.16} & 4.15 \\ \hline
Ours (mean) & 4.90          & 4.34          & 3.09          & 3.16          \\
Ours (GT) & 5.48          & 4.59          & 3.45          & 3.26          \\ \hline
\end{tabular}
\end{center}
\caption{Median scores of SDR for the MUSDB18 dataset.}
\label{table:algorithm}
\end{table}

\subsection{Algorithm Comparison}

In this subsection, we compare our method to other methods with the evaluation result of the MUSDB18 dataset. 
As stated above, our method's output is dependent on the encoded latent vector from a query. For the comparison with other methods that do not require a query, therefore, we used the mean vector in the latent space encoded from the training samples for each source -- \textit{i.e.}, we ended up using four mean latent vectors for `vocals', `drums', `bass', and `other', respectively. Additionally, to show the upper bound of our proposed method, we used the encoded latent vector of the ground truth (GT) signal from test set. Note also that the separation is done with a single network. 

Table \ref{table:algorithm} shows the median scores of SDR of methods reported in SiSEC2018 \cite{stoter20182018}, including our method denoted as Ours. 
Although the proposed algorithm did not achieve the best performance, the results show that it is comparable to the other deep learning-based models that are dedicated to separating just four sources in the dataset. This means that our method is not limited to query-based separation, but also can be used for general music source separation just like as other conventional methods. Additionally, there is room for improvement: applying the multi-channel Wiener filter and/or using other architecture for the separator besides U-net could be such an option.

\section{Conclusion}

In this study, we presented a novel framework, consisting of Query-net and Separator, for audio query-based music source separation. 
Experiment results showed that our method is scalable as the Query-net directly encodes audio query into a latent space.
The latent space is interpretable as was shown by the t-SNE visualization and latent interpolation experiments.
Furthermore, we have introduced various utilities of the proposed framework including manual and automated approach showing the promise of audio-query based source separation.
As a future work, we plan to investigate more adequate conditioning method for audio and better neural architecture for performance improvement.

\section{Acknowledgement}
This work was supported partly by Kakao and Kakao Brain corporations and partly supported by Institute for Information \& Communications Technology Planning \& Evaluation(IITP) grant funded by the Korea government(MSIT) (No.2019-0-01367, Infant-Mimic Neurocognitive Developmental Machine Learning from Interaction Experience with Real World (BabyMind)).
\bibliography{ISMIRtemplate}

%
%
%
%

\end{document}